\documentclass[journal,twoside,web]{ieeecolor}

\usepackage{generic}
\usepackage{cite}
\usepackage{amsmath,amssymb,amsfonts}
\usepackage{graphicx}
\usepackage{textcomp}

\usepackage{graphicx}
\usepackage{subfigure}
\usepackage{epsfig} 
\usepackage{cite}
\usepackage{lcsys}
\usepackage{amsmath,amssymb,amsfonts}
\usepackage{graphicx}
\usepackage{textcomp}
\usepackage{color}
\usepackage{bbm}

\usepackage{graphicx}
\usepackage{mathtools}
\usepackage{multirow}
\usepackage{lineno,hyperref}
\usepackage{url}
\usepackage{cleveref}
\usepackage{subcaption}
\usepackage{soul}
\usepackage{bm}
\usepackage{tikz}
\usepackage{algorithm}
\usepackage{algorithmicx}
\usepackage{algpseudocode}
\usepackage{caption}
\usepackage{mathtools}
\mathtoolsset{showonlyrefs,showmanualtags}

\DeclareMathAlphabet{\pazocal}{OMS}{zplm}{m}{n}

\newcommand{\ve}{\boldsymbol}

\newcommand{\eg}{\emph{e.g.}}
\newcommand{\ie}{\emph{i.e.}}
\newcommand{\cf}{cf.}

\DeclareMathOperator{\diag}{diag}

\renewcommand{\leq}{\leqslant}
\renewcommand{\geq}{\geqslant}
\renewcommand{\epsilon}{\varepsilon}
\newtheorem{defn}{Definition}
\newtheorem{thm}{Theorem}
\newtheorem{lem}{Lemma}

\usepackage{pgfplots}
\DeclareUnicodeCharacter{2212}{−}
\usepgfplotslibrary{groupplots,dateplot}
\usetikzlibrary{patterns,shapes.arrows}

\usepackage{multirow}
\usepackage{color}
\usepackage{paralist}

\pgfplotsset{compat=1.18}

\begin{document}

\def\BibTeX{{\rm B\kern-.05em{\sc i\kern-.025em b}\kern-.08em
    T\kern-.1667em\lower.7ex\hbox{E}\kern-.125emX}}
\markboth{\journalname, VOL. XX, NO. XX, XXXX 2017}
{Author \MakeLowercase{\textit{et al.}}: Preparation of Papers for IEEE Control Systems Letters (August 2022)}

\title{\scalebox{0.91}{Data-Driven Estimation of Structured Singular Values}}

\author{Margarita A. Guerrero, Braghadeesh Lakshminarayanan, and Cristian R. Rojas, \IEEEmembership{Senior Member, IEEE}
\thanks{}
\thanks{This work was partially supported by the Wallenberg AI, Autonomous Systems and Software Program (WASP) funded by the Knut and Alice Wallenberg Foundation. The authors are with the Division of Decision and Control Systems, KTH Royal Institute of Technology, 100 44 Stockholm, Sweden (e-mails: mags3@kth.se, blak@kth.se, crro@kth.se). 
}
\thanks{Code available at \href{https://github.com/mags-ono/SSV}{https://github.com/mags-ono/SSV}.}
}

\maketitle
\thispagestyle{empty}

\begin{abstract}
Estimating the size of the modeling error is crucial for robust control. Over the years, numerous metrics have been developed to quantify the model error in a control relevant manner. One of the most important such metrics is the structured singular value, as it leads to necessary and sufficient conditions for ensuring stability and robustness in feedback control under structured model uncertainty. Although the computation of the structured singular value is often intractable, lower and upper bounds for it can often be obtained if a model of the system is known. In this paper, we introduce a fully data-driven method to estimate a lower bound for the structured singular value, by conducting experiments on the system and applying power iterations to the collected data. Our numerical simulations demonstrate that this method effectively lower bounds the structured singular value, yielding results comparable to the MATLAB$^\copyright$ Robust Control Toolbox.
\end{abstract}

\section{Introduction}

Modeling plays a crucial role in designing feedback controllers. Numerous techniques for modeling dynamical systems have been developed in the literature~\cite{Ljung:99}. Controllers that are based on these models aim to ensure stability of the closed loop, assuming that the model is a faithful representation of the system. However, models often fail to capture every nuance of real-world dynamical systems, as they are inevitably subject to \emph{modeling errors}, which need to be accounted for in control design.

Many robust control theories have been developed to address modeling errors and quantify uncertainties~\cite{zhoudoyle}. One popular approach involves treating the modeling error as a linear operator of bounded $\mathcal{H}_\infty$ norm, which is defined as the supremum over all frequencies of the largest singular value of its frequency response. For linear systems, the $\mathcal{H}_\infty$ norm of the modeling error can be directly estimated from experiments on the system by employing data-driven methods, such as power iterations~\cite{wahlberg2010non,rojas2012analyzing} and Thompson sampling~\cite{muller2019gain}, without the need for a parametric model. Other related concepts are the \emph{input passivity index}, which measures how closely a system is to a passive one, the \emph{$\nu$-gap metric} that measures how far apart two closed loop systems are, and the \emph{optimal stability margin}, which measures the level of unmodelled uncertainty that a closed loop system can handle before it becomes unstable. Data-driven approaches to compute these quantities have been proposed in~\cite{koch2022determining,koenings2017data}.

In the case of structured modeling error, an appropriate metric for the design of robust controllers is the \emph{structured singular value}~\cite{doyle1982analysis}. For systems with unstructured uncertainty, the structured singular value coincides with the $\mathcal{H}_{\infty}$ norm. In the presence of structured uncertainty, however, the structured singular value provides a tighter measure of how large the modeling error can be (in terms of its $\mathcal{H}_\infty$ norm) without leading to instability.


In spite of its usefulness for the establishment of necessary and sufficient conditions for the stability of feedback controllers under structured modeling errors, the structured singular value cannot be easily computed, as its calculation is in general NP-hard~\cite{doyle1982analysis}, so in practice only lower and upper bounds can be determined. 

In~\cite{packard1988power}, a model-based power iteration–based scheme was introduced to estimate a lower bound on the structured singular value of a system. The structured singular value is defined via a minimization over a set of block-diagonal matrices that model structured uncertainties, encompassing both repeated scalar blocks and full blocks. The authors derived an iterative algorithm, drawing inspiration from power iterations, which alternates between updating unitary matrices and diagonal scaling matrices until convergence to an equilibrium point is attained. At convergence, the algorithm provides a valid lower bound for the structured singular value. Recent works propose tighter structured singular value lower and upper bounds for special block structures~\cite{musthaqcomplex,MUSHTAQ2024111717,prajapati2024structuredsingularvaluesapplication}; however they also require a model for the system and impose additional constraints on structural uncertainties. 

In this paper, we introduce a fully data-driven power iteration scheme~\cite{rojas2012analyzing} to compute a lower bound on the structured singular value of a linear dynamical system. Our method overcomes the need for an explicit model of the system, by instead relying on experimental data. While upper bounds aid robustness certification, estimating a lower bound offers a conservative robustness metric—small values imply the system tolerates significant structured uncertainty without instability. To the best of our knowledge, our paper proposes the first fully data-driven approach to compute such lower bounds without additional assumptions on the uncertainties, unlike recent works~\cite{musthaqcomplex,MUSHTAQ2024111717,prajapati2024structuredsingularvaluesapplication}. Specifically, our main contributions can be summarized as follows:
\begin{itemize}
    \item We propose a data-driven approach to numerically compute a lower bound on the structured singular value for dynamical systems.
    \item We demonstrate our method’s effectiveness through numerical simulations, where the estimated lower bounds agree with MATLAB$^\copyright$ \texttt{mussv} in most cases.
\end{itemize}

The remainder of this paper is organized as follows: In Section~\ref{sec: setup}, we state our problem setup. Section~\ref{sec: prelims} provides some preliminaries on the power method. In Section~\ref{sec: approach}, we outline our proposed approach, and demonstrate its efficacy in Section~\ref{sec: Simulation}. Finally, the paper is concluded in Section~\ref{sec: conclusion}.

\noindent \textbf{Notation.} Vectors and matrices are written in bold. If $\ve{M} \in \mathbb{C}^{n \times n}$ is a complex matrix, $\bar{\sigma}(\ve{M})$ denotes its largest singular value, and $\rho(\ve{M})$ its spectral radius. The symbol $\diag( \cdots )$ denotes a block-diagonal matrix with blocks given by its arguments. The vector $\ve{e}_p$ represents the $p$-th canonical basis vector in $\mathbb{R}^n$. The set $\mathbb{S}^m$ consists of all Hermitian positive definite matrices of size $m \times m$. The symbols $q$ and $z$ denote the forward shift operator and the complex frequency variable in the $z$-domain, respectively.


\section{Problem Statement} \label{sec: setup}

Consider a linear time-invariant square multivariable discrete-time dynamical system defined by its transfer function $\ve{G}(z)$. We assume that it is composed of a ``nominal'' stable and strictly proper model $\ve{G}_0(z)$ and a stable block $\ve{\Delta}(z)$ denoting a multiplicative uncertainty. Let $\ve{U}(z) \in \mathbb{C}^n$ and $\ve{Y}(z) \in \mathbb{C}^n$ denote the Z-transforms of the input and output signals of $\ve{G}$, respectively. Then,  
\begin{align*}
\ve{Y}(z) = \underbrace{[\ve{I} + \ve{G}_0(z) \ve{\Delta}(z)]^{-1} \ve{G}_0(z)}_{=:\ve{G}(z)} \ve{U}(z).
\end{align*}
Since $\ve{G}_0$ is stable and strictly proper (\ie, the poles of $\det[\ve{G}_0(z)]$ are in the open unit disk $\mathbb{D} := \{ z \in \mathbb{C}\colon |z| < 1 \}$), the full system $\ve{G}$ is stable if and only if all the zeros of $\det[\ve{I} + \ve{G}_0(z) \ve{\Delta}(z)]$ are in the open unit disk. Furthermore, since $\ve{\Delta}$ is also stable, the poles of $\det[\ve{I} + \ve{G}_0(z) \ve{\Delta}(z)]$ are all in $\mathbb{D}$, so $\ve{G}$ is stable if and only if, by the Nyquist criterion, the curve that $z \mapsto \det[\ve{I} + \ve{G}_0(z) \ve{\Delta}(z)]$ draws on the complex plane, as $z$ runs over $\mathbb{T} := \partial \mathbb{D}$ counter-clockwise, does \emph{not} encircle (clockwise) the point $0$.

Suppose that $\ve{\Delta}$ is known to be of the form $\diag(\delta_1 \ve{I}_{r_1}, \dots, \delta_s \ve{I}_{r_s}, \ve{\Delta}_1, \dots, \ve{\Delta}_f)$, where $\delta_1, \dots, \delta_s \in \mathbb{C}$ and $\ve{\Delta}_1, \dots, \ve{\Delta}_f$ are stable dynamical systems of sizes $m_1 \times m_1$, \dots, $m_f \times m_f$, respectively. We say that $\ve{\Delta} \in \Delta$ if it possesses this structure. Then, the \emph{structured singular value} of $\ve{G}_0$ is defined as 

\vspace{-1 em}
{\fontsize{9}{10}\selectfont
\begin{equation}
\mu_\Delta(\mathbf{G}_0) :=
\frac{1}{\min\Biggl\{ \|\mathbf{\Delta}\|_\infty \colon 
\begin{array}{c}
\ve{\Delta} \in \Delta, \\[0.3em]
\det\Bigl[\mathbf{I} + \mathbf{G}_0(z)\,\mathbf{\Delta}(z)\Bigr] = 0 \\[0.3em]
\text{for some } z \in \mathbb{T}
\end{array}
\Biggr\}}.
\end{equation}}
The interest in $\mu_\Delta(\mathbf{G}_0)$ lies in its use to characterize the stability of $\ve{G}$, as stated in the following theorem. 

\begin{thm}[\hspace{-0.2mm}{\cite[Theorem~11.8]{zhoudoyle}}]
The system $\ve{G} = [\ve{I} + \ve{G}_0 \ve{\Delta}]^{-1} \ve{G}_0$ is stable for all $\ve{\Delta} \in \Delta$ such that $\| \ve{\Delta} \|_\infty \leq \alpha$ if and only if $\alpha\, \mu_\Delta(\ve{G}_0) < 1$.
\end{thm}

This result is an extension of the \emph{small gain theorem} to structured uncertainty, and it is fundamental for the study of the stability of uncertain systems in robust control~\cite{zhoudoyle}. The quantity $\mu_\Delta(\ve{G}_0)$ can be evaluated frequency-wisely as

\vspace{-0.9 em}
\begingroup
\fontsize{8}{10}\selectfont
\begin{equation}
\begin{aligned}
\mu_\Delta(\ve{G}_0)
:= \sup_{\omega \in (-\pi, \pi]} \frac{1}{\min\{ \bar{\sigma}(\ve{\Delta})\colon \ve{\Delta} \in \Delta, \det[\ve{I} + \ve{G}_0(e^{i \omega}) \ve{\Delta}] = 0 \}},
\end{aligned}
\end{equation}
\endgroup
where now $\ve{\Delta}$ is a (static) complex matrix having the structure given by $\Delta$.

Unfortunately, computing $\mu_\Delta(\ve{G}_0)$ is hard, hence practical approaches typically rely on lower and upper bounds~\cite{doyle1982analysis}.
An appealing approach that exists in the literature~\cite{packard1988power} is to compute a lower bound on $\mu_\Delta(\ve{G}_0)$ based on the power iterations method~\cite{golub2013matrix} and a model for $\ve{G}_0$. Inspired by that method, our main contribution in this paper is to propose a fully data-driven approach to compute a lower bound on $\mu_\Delta(\ve{G}_0)$ that does not require knowledge of $\ve{G}_0$. To that end, we consider a discrete-time LTI system described by
\begin{align}
\ve{y}_k = \ve{G}_0(q) \ve{u}_k + \ve{e}_k,
\end{align}
where $\ve{G}_0(q)$ is an unknown but stable system, $\ve{u}_k$ is the input at time $k$, $\ve{y}_k$ is the corresponding output, and $\ve{e}_k$ is a zero-mean additive disturbance. 
Then, the underlying data-generation process follows an iterative redesign of the input signal $\{\ve{u}_k\}_{k=1}^N$, based on the power iteration method, to focus sampling on frequencies that contribute most to $\mu_\Delta(\ve{G}_0)$. This redesign uses information extracted from the output $\{\ve{y}_k\}_{k=1}^N$ obtained in previous iterations. We assume $\ve{e}_k = \ve{0}$ throughout the analysis, while noise is included in the simulations of Section~\ref{sec: Simulation} to evaluate robustness. To ensure consistency with the power method in~\cite{packard1988power}, we have adopted the notation used in that paper.

Before we propose our approach, we shall present several technical preliminaries that are required for this paper.

\section{Preliminaries} \label{sec: prelims}
In this section, we review the model-based power method for computing a lower bound on $\mu_\Delta$. The results stated in this section appear in~\cite{packard1988power,doyle1982analysis}. 

Let us first define the following structured block-diagonal uncertainty $\Delta$ and bounded uncertainty $\text{B}\Delta$ sets as
\begin{align*}
\Delta &:= \{ \diag(\delta_1 \ve{I}_{r_1}, \dots, \delta_s \ve{I}_{r_s}, \ve{\Delta}_1, \dots, \ve{\Delta}_f)\colon\\
&\quad \delta_1, \dots, \delta_s \in \mathbb{C},\, \ve{\Delta}_1 \in \mathbb{C}^{m_1 \times m_1}, \dots, \ve{\Delta}_f \in \mathbb{C}^{m_f \times m_f}\}, \\
\text{B} \Delta &:= \{ \ve{\Delta} \in \Delta\colon \bar{\sigma}(\ve{\Delta}) \leq 1 \},
\end{align*}
where $r_1, \dots, r_s, m_1, \dots, m_f$ are fixed positive integers such that $r_1 + \cdots + r_s + m_1 + \cdots + m_f = n$.
Observe that $\Delta$ is a (complex) linear subspace of $\mathbb{C}^{n \times n}$. Based on these sets, the structured singular value can be defined at a given frequency $\omega \in [-\pi, \pi)$ (by letting $\ve{M} = \ve{G}_0(e^{i \omega})$) as


\begin{defn}[Structured singular value]
For $\ve{M} \in \mathbb{C}^{n \times n}$, let

\vspace{-1.1 em}
{\fontsize{8.5}{9.5}\selectfont
\begin{align*}
\mu'_\Delta(\ve{M}) := \frac{1}{\min\ \{ \bar{\sigma}(\ve{\Delta})\colon \ve{\Delta} \in \Delta,\, \det(\ve{I} + \ve{M} \ve{\Delta}) = 0 \} }.
\end{align*}}
In case there is no $\ve{\Delta} \in \Delta$ for which $\det(\ve{I} + \ve{M} \ve{\Delta}) = 0$, we define $\mu'_\Delta(\ve{M}) = 0$. Note that $\mu_{\Delta}(G_{0})=\sup_{\omega}\mu_{\Delta}^{\prime}(M).$
\end{defn}
Let us define the sets of unitary and structured Hermitian positive definite matrices $\pazocal{Q}$ and $\pazocal{D}$, respectively, as:
\begin{equation*}
\begin{split}
\pazocal{Q} &:= \{ \ve{\Delta} \in \Delta\colon \ve{\Delta}^H \ve{\Delta} = \ve{I} \}, \\
\pazocal{D} &:= \{ \diag(\ve{D}_1, \dots, \ve{D}_s, d_1 \ve{I}_{m_1}, \dots, d_1 \ve{I}_{m_f})\colon\\
&\quad \qquad \ve{D}_1 \in \mathbb{S}^{r_1}, \dots, \ve{D}_s \in \mathbb{S}^{r_s},\ d_1, \dots, d_f \in \mathbb{R}_+ \}.
\end{split}
\end{equation*}
These sets satisfy the following properties:
\begin{lem}[\hspace{-0.2mm}{\cite[p.1]{packard1988power}}]  \label{lem:properties}
Let $\ve{\Delta} \in \Delta$, $\ve{Q} \in \pazocal{Q}$ and $\ve{D} \in \pazocal{D}$. Then, $\ve{Q}^H \in \pazocal{Q}$, $\ve{Q} \ve{\Delta} \in \Delta$, $\ve{\Delta} \ve{Q} \in \Delta$, $\bar{\sigma}(\ve{Q} \ve{\Delta}) = \bar{\sigma}(\ve{\Delta} \ve{Q}) = \bar{\sigma}(\ve{\Delta})$, and $\ve{D} \ve{\Delta} = \ve{\Delta} \ve{D}$.
%
\end{lem}

The following theorem establishes bounds on $\mu'_{\Delta}$. 

\begin{thm}[\hspace{-0.2mm}{\cite[Theorem~1]{doyle1982analysis}}] \label{thm:bounds_mu}
For all $\ve{M} \in \mathbb{C}^{n \times n}$,

{\fontsize{9}{10}\selectfont
\vspace{0.1 em}
\begin{center}
\scalebox{0.96}{ 
$\displaystyle
\max_{\ve{Q} \in \pazocal{Q}} \rho(\ve{Q} \ve{M})
= \max_{\ve{\Delta} \in \text{B}\Delta} \rho(\ve{\Delta} \ve{M})
= \mu'_\Delta(\ve{M})
\leq \inf_{\ve{D} \in \pazocal{D}} \bar{\sigma}(\ve{D} \ve{M} \ve{D}^{-1})
$.}
\end{center}}
\end{thm}
\vspace{0.3 em}


As established in Theorem~\ref{thm:bounds_mu}, our goal is to derive a method for finding a local maximum of the function $\ve{\Delta} \mapsto \rho(\ve{\Delta} \ve{M})$ over all $\ve{\Delta} \in \text{B}\Delta$; every such local maximum provides a lower bound to $\mu'_\Delta(\ve{M})$. Packard et al. derived in \cite{packard1988power} the following necessary characterization of $\mu'_\Delta(\ve{M})$, based on Theorem~\ref{thm:bounds_mu}.

\begin{thm}[\hspace{-0.2mm}{\cite[Theorem~6.3]{packard1988power}}] \label{thm:optim_cond}
Given a matrix $\ve{M} \in \mathbb{C}^{n \times n}$, let $\ve{Q}_0 \in \pazocal{Q}$ achieve the global maximum in $\max_{\ve{Q} \in \pazocal{Q}} \rho(\ve{Q} \ve{M})$, and assume that the maximum eigenvalue of $\ve{Q}_0 \ve{M}$ is simple, real and positive; call it $\mu$. If $\ve{x} = [\ve{x}_{r_1}^T, \dots, \ve{x}_{r_s}^T, \ve{x}_{m_1}^T, \dots, \ve{x}_{m_f}^T]^T \in \mathbb{C}^n$ and $\ve{y} = [\ve{y}_{r_1}^T, \dots, \ve{y}_{r_s}^T, \ve{y}_{m_1}^T, \dots, \ve{y}_{m_f}^T]^T\in \mathbb{C}^n$ are right and left eigenvectors of $\ve{Q}_0 \ve{M}$ associated with this eigenvalue (where the partitions of $\ve{x}$ and $\ve{y}$ are compatible with the block structure of $\Delta$) such that $\ve{y}^H \ve{x} = 1$, $\ve{y}_{r_j}^H \ve{x}_{r_j} \neq 0$ for all $j = 1, \dots, s$, and $\ve{x}_{m_k}, \ve{y}_{m_k} \neq 0$ for all $k = 1, \dots, f$, 
%
%
then there exists a $\ve{D} \in \pazocal{D}$ such that
%

\vspace{-2.7 em}
{\fontsize{10}{11.5}\selectfont
\[
\ve{Q}_0 \ve{D} \ve{M} \ve{D}^{-1} (\ve{D} \ve{x}) = \mu \ve{D} \ve{x}, \,
\ve{x}^H \ve{D} \ve{Q}_0 \ve{D} \ve{M} \ve{D}^{-1} = \mu \ve{x}^H \ve{D}.\vspace{-0.55 em} 
\]}
\end{thm}

\smallskip
According to Theorem~\ref{thm:optim_cond}, and via the change of variable $\bar{\ve{x}} := \ve{D} \ve{x}$, given a matrix $\ve{M} \in \mathbb{C}^{n \times n}$, to find $\mu = \max_{\ve{Q} \in \pazocal{Q}} \rho(\ve{Q} \ve{M})$ we can try to find matrices $\ve{Q} \in \pazocal{Q}$ and $\ve{D} \in \pazocal{D}$, and a vector $\bar{\ve{x}} \in \mathbb{C}^n$ with $\| \bar{\ve{x}} \| = 1$ such that
\vspace{-1.4 em}
\[
\ve{Q} \ve{D} \ve{M} \ve{D}^{-1} \bar{\ve{x}} = \mu \bar{\ve{x}} , \,
\ve{D}^{-1}  \ve{M}^H \ve{D} \ve{Q}^H \bar{\ve{x}}  = \mu \bar{\ve{x}} ,\vspace{-1.5 em} 
\]
which can, in turn, be re-written as
\vspace{-1.4 em}
\[\ve{M} (\ve{D}^{-1} \bar{\ve{x}} ) = \mu (\ve{D}^{-1} \ve{Q}^H \bar{\ve{x}} ) , \, 
\ve{M}^H (\ve{D} \ve{Q}^H \bar{\ve{x}} ) = \mu (\ve{D}\bar{\ve{x}} ).\vspace{-1.4 em}
\]
For fixed $\ve{Q} \in \pazocal{Q}$ and $\ve{D} \in \pazocal{D}$, let us define the vectors
%
\vspace{-1.3 em}
\[
\ve{a}\coloneqq \ve{D}^{-1} \ve{Q}^H \bar{\ve{x}} , 
\ve{b}\coloneqq \ve{D}^{-1} \bar{\ve{x}} ,
\ve{w}\coloneqq \ve{D} \bar{\ve{x}} , \text{ and }
\ve{z}\coloneqq \ve{D} \ve{Q}^H \bar{\ve{x}} .\vspace{-1.3 em}
\]
With these definitions, we have that $\ve{M} \ve{b} = \mu \ve{a}$ and $\ve{M}^H \ve{z} = \mu \ve{w}$. We can eliminate $\bar{\ve{x}} $ from these definitions, using Lemma~\ref{lem:properties} and replacing $\ve{D}^2$ with $\ve{D}$, obtaining
%
\vspace{-1.4 em}
\[
\ve{b} = \ve{Q} \ve{a},\;
\ve{b} = \ve{D}^{-1} \ve{w},\;
\ve{z} = \ve{D} \ve{a},\;
\ve{z} = \ve{Q}^H \ve{w}.\vspace{-1.65 em} 
\]

After eliminating $\bar{\ve{x}} $, we want to remove $\ve{Q}$ and $\ve{D}$ from these definitions. To this end, let $\ve{a} = [\ve{a}_{r_1}^T, \dots, \ve{a}_{r_s}^T, \ve{a}_{m_1}^T, \dots, \ve{a}_{m_f}^T]^T \in \mathbb{C}^n$ according to the structure of $\Delta$, and similarly for the other vectors. Then, we have the following lemma.

\begin{lem} [\hspace{-0.2mm}{\cite[Lemma~7.1]{packard1988power}}] \label{lem:elimination}
Given non-zero vectors $\ve{a}, \ve{b}, \ve{w}, \ve{z} \in \mathbb{C}^n$, there are matrices $\ve{Q} \in \pazocal{Q}$ and $\ve{D} \in \pazocal{D}$ such that $\ve{b} = \ve{Q} \ve{a}$, $\ve{b} = \ve{D}^{-1} \ve{w}$, $\ve{z} = \ve{D} \ve{a}$, and $\ve{z} = \ve{Q}^H \ve{w}$, if and only if
%

\vspace{-2.3 em}
{\fontsize{8.5}{9.5}\selectfont
\begin{align*}
&\ve{z}_{r_j} = \frac{\ve{w}_{r_j}^H \ve{a}_{r_j}}{|\ve{w}_{r_j}^H \ve{a}_{r_j}|} \ve{w}_{r_j}; \,\,
\ve{b}_{r_j} = \frac{\ve{a}_{r_j}^H \ve{w}_{r_j}}{|\ve{a}_{r_j}^H \ve{w}_{r_j}|} \ve{a}_{r_j}, \quad j = 1, \dots, s, \\
&\ve{z}_{m_k} = \frac{\| \ve{w}_{m_k} \|}{\| \ve{a}_{m_k} \|} \ve{a}_{m_k}; \, \, \ve{b}_{m_k} = \frac{\| \ve{a}_{m_k} \|}{\| \ve{w}_{m_k} \|} \ve{w}_{m_k}, \quad k = 1, \dots, f.
\end{align*}}
\end{lem}
\vspace{0.4 em}
Combining the conditions in Lemma~\ref{lem:elimination} with the conditions that $\ve{M} \ve{b} = \mu \ve{a}$ and $\ve{M}^H \ve{z} = \mu \ve{w}$ suggests the following power method for determining a lower bound on $\mu$:

{\fontsize{8.5}{9.5}\selectfont
\begin{align*}
\tilde{\mu}(l+1) &\gets \| \ve{M} \ve{b}(l) \|, \, 
\ve{a}(l+1) \gets \frac{1}{\tilde{\mu}(l+1)} \ve{M} \ve{b}(l) \\
\ve{z}_{r_j}(l+1) &\gets \frac{\ve{w}_{r_j}^H(l) \ve{a}_{r_j}(l+1)}{|\ve{w}_{r_j}^H(l) \ve{a}_{r_j}(l+1)|} \ve{w}_{r_j}(l), \quad j = 1, \dots, s \\
\ve{z}_{m_k}(l+1) &\gets \frac{\| \ve{w}_{m_k}(l) \|}{\| \ve{a}_{m_k}(l+1) \|} \ve{a}_{m_k}(l+1), \quad k = 1, \dots, f \\
\bar{\mu}(l+1) &\gets \| \ve{M}^H \ve{z}(l+1) \| \\
\ve{w}(l+1) &\gets \frac{1}{\bar{\mu}(l+1)} \ve{M}^H \ve{z}(l+1) \\
\ve{b}_{r_j}(l+1) &\gets \frac{\ve{a}_{r_j}^H(l+1) \ve{w}_{r_j}(l+1)}{|\ve{a}_{r_j}^H(l+1) \ve{w}_{r_j}(l+1)|} \ve{a}_{r_j}(l+1), \quad j = 1, \dots, s \\
\ve{b}_{m_k}(l+1) &\gets \frac{\| \ve{a}_{m_k}(l+1) \|}{\| \ve{w}_{m_k}(l+1) \|} \ve{w}_{m_k}(l+1), \quad k = 1, \dots, f.
\end{align*}}
Here the entire set of assignments is iterated over $l=0, 1, \dots$ until convergence, where the lower bound on $\mu$ is obtained as an equilibrium point $\tilde{\mu}(l) = \bar{\mu}(l)$. Here $\ve{b}(0)$ and $\ve{w}(0)$ are chosen as unit vectors. 

\section{Proposed Approach} \label{sec: approach}
In this section, we propose a data-driven approach to compute a lower bound on $\mu_\Delta$. Assuming that a model of the plant is unavailable and only time-domain experiments can be conducted, we adapt the frequency-domain power iteration scheme of~\cite{packard1988power} to operate purely on input-output data.

Firstly, note that the power iteration of Section~\ref{sec: prelims} is applied separately at each frequency. Since experiments are conducted in the time domain, it is necessary to map signals to the frequency domain, where the updates of $\ve{b}$, $\ve{a}$, $\ve{w}$, and $\ve{z}$—\cf~Lemma~\ref{lem:elimination}—are computed over the relevant frequency range. To that end, we associate with each time-domain signal its discrete Fourier transform (DFT), and use its inverse to reconstruct the time-domain signals when needed.

In the sequel, we add the frequency argument $m$ to vectors $\ve{b}$, $\ve{a}$, $\ve{w}$, and $\ve{z}$ and write them in uppercase to denote their frequency-domain versions, with lowercase denoting their time-domain counterparts. This leads to the following frequency-domain power iteration scheme (for each $m = 0,\ldots,N{-}1$; $j = 1,\ldots,s$; and $k = 1,\ldots,f$), where we have replaced $\ve{M}$ by $\ve{G}_0(e^{i 2 \pi m / N})$, which will be reinterpreted through experimental data in the next step:
%
%

{\fontsize{8.5}{9.5}\selectfont
\begin{IEEEeqnarray}{rCl}
\tilde{\mu}[l+1,m] &\gets& \left\| \ve{G}_0(e^{i 2 \pi m / N}) \ve{B}[l,m] \right\| \label{eq:mu_update} \\
\ve{A}[l+1,m] &\gets& \frac{1}{\tilde{\mu}[l+1,m]} \ve{G}_0(e^{i 2 \pi m / N}) \ve{B}[l,m] \label{eq:A_update} \\
\ve{Z}_{r_j}[l+1,m] &\gets& \frac{\ve{W}_{r_j}^H[l,m] \ve{A}_{r_j}[l+1,m]}{\left| \ve{W}_{r_j}^H[l,m] \ve{A}_{r_j}[l+1,m] \right|} \ve{W}_{r_j}[l,m] \label{eq:Zr_update} \\
\ve{Z}_{m_k}[l+1,m] &\gets& \frac{\left\| \ve{W}_{m_k}[l,m] \right\|}{\left\| \ve{A}_{m_k}[l+1,m] \right\|} \ve{A}_{m_k}[l+1,m] \label{eq:Zm_update} \\
\bar{\mu}[l+1,m] &\gets& \left\| \ve{G}_0^T(e^{-i 2 \pi m / N}) \ve{Z}[l+1,m] \right\| \label{eq:mubar_update} \\
\ve{W}[l+1,m] &\gets& \frac{1}{\bar{\mu}[l+1,m]} \ve{G}_0^T(e^{-i 2 \pi m / N}) \ve{Z}[l+1,m] \label{eq:W_update} \\
\ve{B}_{r_j}[l+1,m] &\gets&
\scalebox{0.91}{$\displaystyle
\frac{\ve{A}_{r_j}^H[l+1,m] \ve{W}_{r_j}[l+1,m]}{\left| \ve{A}_{r_j}^H[l+1,m] \ve{W}_{r_j}[l+1,m] \right|} \ve{A}_{r_j}[l+1,m]
$}
\label{eq:Br_update}
\\
\ve{B}_{m_k}[l+1,m] &\gets& \frac{\left\| \ve{A}_{m_k}[l+1,m] \right\|}{\left\| \ve{W}_{m_k}[l+1,m] \right\|} \ve{W}_{m_k}[l+1,m] \label{eq:Bm_update}
\end{IEEEeqnarray}
}
To implement this algorithm in a data-driven setting, we must re-write the operations involving $\ve{G}_0(e^{i 2 \pi m / N})$ and $\ve{G}_0^T(e^{-i 2 \pi m / N})$ in terms of time-domain experiments. Specifically, we define $\ve{P}[l,m] := \ve{G}_0(e^{i 2 \pi m / N}) \ve{B}[l,m]$ and $\ve{R}[l+1,m] := \ve{G}_0^T(e^{-i 2 \pi m / N}) \ve{Z}[l+1,m]$, which must now be approximated from data. The computation of $\ve{P}$ can be carried out as follows:
%

\vspace{-1 em}
{\fontsize{8.5}{9.5}\selectfont
\begin{IEEEeqnarray}{rCl}
\ve{b}[l,t] & \gets & \frac{1}{N} \sum_{m=0}^{N-1} \ve{B}[l,m] e^{i 2 \pi m t / N}, \quad t = 1, \dots, N \label{eq:b_update} \\
\ve{p}[l,:] & \gets & \ve{G}_0(q) \ve{b}[l,:] \label{eq:p_update} \\
\ve{P}[l,m] & \gets & \sum_{t=1}^N \ve{p}[l,t] e^{-i 2 \pi m t / N}, \quad m = 0, 1, \dots, N-1 \label{eq:P_update}
\end{IEEEeqnarray}
}
\vspace{-0.8 em}

Here, $\ve{p}[l,:]$ is the output of $\ve{G}_0$ when excited with $\ve{b}[l,t]$, a time-domain signal derived from the frequency-domain vector $\ve{B}[l,:]$ in Lemma~\ref{lem:elimination}, chosen to span the desired frequencies. Similarly, $\ve{r}[l+1,:]$ is the response of $\ve{G}_0^T$ to $\ve{z}[l+1,:]$, also from Lemma~\ref{lem:elimination} and mapped to time, since all experiments on $\ve{G}_0$ are conducted in the time domain.


\vspace{0.1em}
The computation of $\ve{R}$ is a bit trickier, due to the transpose and complex conjugate operations. The complex conjugate operation in the frequency domain corresponds to applying $\ve{G}_0$ ``backwards in time'', or by replacing $t$ with $N+1 - t$ in the computation of the discrete Fourier transform and its inverse. To account for the transpose of $\ve{G}_0$, we can appeal to the trick in \cite[Eq.~(21)]{oomen2014iterative}, according to which

\vspace{-0.9 em}
{\fontsize{8.5}{9.5}\selectfont
\begin{IEEEeqnarray}{rCl}
\ve{G}_0^T(q) & = & \sum_{\alpha=1}^n \sum_{\beta=1}^n \ve{e}_\alpha \ve{e}_\beta^T \ve{G}_0(q) \ve{e}_\alpha \ve{e}_\beta^T. \label{eq:G0_transposed}\vspace{-0.3 em}
\end{IEEEeqnarray}
}
Combining these ideas, we obtain the following pseudo code for computing $\ve{R}$:
%
%

\vspace{-1 em}
{\fontsize{8.5}{9.5}\selectfont
\begin{IEEEeqnarray}{ll}
\ve{z}[l+1,t] \gets \frac{1}{N} \sum_{m=0}^{N-1} \ve{Z}[l+1,m] e^{i 2 \pi m (1-t) / N},\,\, \scalebox{0.92}{$t = 1, \dots, N$} \label{eq:z_update} \\
\ve{r}[l+1,t] \gets \ve{0} \label{eq:r_zero} \\
\IEEEeqnarraymulticol{2}{l}{\hspace{1em}\textbf{for } \ensuremath{\alpha = 1, \dots, n}\textbf{:}} \nonumber\\
\IEEEeqnarraymulticol{2}{l}{\hspace{2em}\textbf{for } \ensuremath{\beta = 1, \dots, n}\textbf{:}} \nonumber\\
\qquad \quad \ve{r}[l+1,:] \gets \ve{r}[l+1,:] + \ve{e}_\alpha \ve{e}_\beta^T \ve{G}_0(q) \ve{e}_\alpha \ve{e}_\beta^T \ve{z}[l+1,:] \label{eq:r_update} \\
\ve{R}[l+1,m] \gets \sum_{t=1}^N \ve{r}[l+1,t] e^{i 2 \pi m (t-1) / N},\,\, \scalebox{0.91}{$m = 0, \dots, N-1$} \label{eq:R_update}\vspace{-0.9 em}
\end{IEEEeqnarray}
}
\begin{algorithm}[b!]
\footnotesize
\caption{Computation of a lower bound on $\mu$}
\label{alg:power_method}
\begin{algorithmic}[1]
\Require {$\ve{B}[0,:]$, $\ve{W}[0,:]$}
\For{$l = 0, 1, \ldots$}
\State \textbf{for} $t = 1, \dots, N,$ \textbf{compute $\ve{b}[l,t]$ via \eqref{eq:b_update}}, \textbf{end for}

\State \textbf{apply $\ve{b}[l,:]$ to $\ve{G}_0(q)$ and observe $\ve{p}[l,:]$ via \eqref{eq:p_update}} 
\State \textbf{for} $m = 0, \dots, N{-}1$, \textbf{compute $\ve{P}[l,m]$ via \eqref{eq:P_update}}, \textbf{end for}
\For{$m = 0, \dots, N-1$}
\State \textbf{compute $\tilde{\mu}[l+1,m]$ via \eqref{eq:mu_update}}
\State \textbf{compute $\displaystyle \ve{A}[l+1,m]$ via \eqref{eq:A_update}}
\State \textbf{for} $j = 1, \dots, s,$ \textbf{compute $\displaystyle \ve{Z}_{r_j}[l+1,m]$ via \eqref{eq:Zr_update}, \textbf{end for}} 
\State \textbf{for} $k = 1, \dots, f,$ \textbf{compute $\displaystyle \ve{Z}_{m_k}[l+1,m]$ via \eqref{eq:Zm_update}, \textbf{end for}} 
\EndFor

\State \textbf{for} $t = 1, \dots, N,$ \textbf{compute $\ve{z}[l+1,t]$ via \eqref{eq:z_update}}, \textbf{end for}
\State \textbf{initialize $\ve{r}[l+1,t]$ via \eqref{eq:r_zero}} 

\For{$\alpha = 1, \dots, n$}
\State \textbf{for} $\beta = 1, \dots, n$ \textbf{apply} $\ve{z}[l+1,:]$ \textbf{to} \eqref{eq:G0_transposed} \textbf{and observe} $\ve{r}[l+1,:]$ \par
\hspace*{0.8em} \textbf{via} \eqref{eq:r_update}, \textbf{end for}

\EndFor
\State \textbf{for} $m = 0, \dots, N{-}1$ \textbf{compute} $\ve{R}[l+1,m]$ via \eqref{eq:R_update}, \textbf{end for}


\For{$m = 0, \dots, N-1$}
\State \textbf{compute $\bar{\mu}[l+1,m]$ via \eqref{eq:mubar_update}}
\State \textbf{compute $\ve{W}[l+1,m]$ via \eqref{eq:W_update}}
\State \textbf{for} $j = 1, \dots, s,$ \textbf{compute $\displaystyle \ve{B}_{r_j}[l+1,m]$ via \eqref{eq:Br_update}}, \textbf{end for} 
\State \textbf{for} $k = 1, \dots, f,$ \textbf{compute $\displaystyle \ve{B}_{m_k}[l+1,m]$ via \eqref{eq:Bm_update}}, \textbf{end for} 

\EndFor

\State \scalebox{0.9}{$\displaystyle \ve{B}[l+1,:] \leftarrow \frac{1}{\sqrt{ \sum_{i=0}^{N-1}\| \ve{B}[l+1,i] \|_2^2  }} \ve{B}[l+1,:]$}
\EndFor
\end{algorithmic}
\end{algorithm}

%
%
Note that the first and last lines do \emph{not} correspond to the standard discrete Fourier transform and its inverse, but to their ``time-reversed'' versions.
Additionally, to prevent signals such as $\ve{b}[l,t]$ from growing unbounded as $l$ increases, we normalize them after each iteration by their $2$-norms. 

The previous discussions finally lead to the pseudo-code for the power method shown in Algorithm~\ref{alg:power_method}. The algorithm terminates when $\bar{\mu} \approx \tilde{\mu}$ and their values remain unchanged across iterations for each frequency, i.e., $\tilde{\mu}(l+1,m) = \tilde{\mu}(l,m)$ and $\bar{\mu}(l+1,m) = \bar{\mu}(l,m)$. Finally, $\mu_{\Delta}(\ve{G}_0)$ is obtained by selecting the maximum $\mu$ across all frequencies.

\vspace{0.4em}
\section{Experiments} \label{sec: Simulation}
This section presents a comprehensive set of simulations to assess the performance of Algorithm~\ref{alg:power_method} against the lower bound provided by the \texttt{mussv} command from the MATLAB$^\copyright$ Robust Control Toolbox \cite[Ch.~10]{Gu2013Robust}.
For clarity, we refer to this lower bound as $\mu_M$ throughout.

\subsection{Experimental Setting}
As described in Algorithm~\ref{alg:power_method}, the computation of the structured singular value requires two experiments on $\ve{G}_0$ and $\ve{G}_0^H$ (the latter being based on \eqref{eq:G0_transposed}). However, to properly account for the real-valued nature of the pulse response of $\ve{G}_0$, and as indicated in \cite[Procedure~2]{oomen2014iterative}, the input signals used in these experiments must be real. This constraint imposes a frequency symmetry condition on the initial vectors $\ve{B}$ and $\ve{W}$, which are otherwise randomly chosen. Due to this randomness, the initial input is a white noise signal, exciting all frequencies uniformly, while subsequent iterations redesign it to focus sampling effort on frequencies where $\mu_\Delta(\ve{G}_0)$ is expected to be maximal. Moreover, to avoid degeneracy, the initial vectors $\ve{b}[l,t]$ and $\ve{z}[l+1,t]$ must have nonzero components along the dominant eigenvector directions; otherwise, the algorithm is heuristically restarted with a different initial condition, i.e., by reinitializing $\ve{B}[0,:]$ and $\ve{W}[0,:]$. Therefore, under these necessary conditions—namely, non-degenerate initialization, spectral symmetry, and a real and simple dominant eigenvalue of $\ve{Q}_0 \ve{M}$, as discussed in Theorem~\ref{thm:optim_cond}—the iterative scheme in Algorithm~\ref{alg:power_method} is expected to converge reliably in many practical scenarios.

Another critical aspect concerns transient effects in physical systems or finite-time simulations. Since the procedure is carried out in the frequency domain, a sufficiently large number of time samples or frequency points must be considered, to reduce the effects of transients. This requirement aligns with the condition $ N \to \infty $ stated in~\cite{oomen2014iterative}, which, in turn, ensures consistency with the requirements of the \textit{power method}~\cite{golub2013matrix}, namely,  $\|\ve{\Delta}\|_{i2} \to \|\ve{\Delta}\|_{\infty}$  as $N \to \infty$.


To clarify the results presented in the next subsection, we denote the uncertainty structure by
\(
\ve{r} = [\,s,\,r_1,\dots,r_s\,],
\,
\mathbf{m} = [\,f,\,m_1,\dots,m_f\,],
\)
where $s$ and $f$ are the numbers of scalar and full blocks, of sizes $r_j$ and $m_k$, respectively. For example,
\(
\mathbf{r} = [2,\,1,\,1], 
\,
\mathbf{m} = [1,\,2]
\)
describes two $1\times1$ scalar blocks and one $2\times2$ full block.



\subsection{Experimental Results}
The experimental evaluation consists of two test cases aimed at assessing the accuracy of the data-driven estimation of the lower bound on the structured singular value and the influence of the uncertainty structure $\ve{\Delta}$ on its performance.

\subsection*{ Test $\#$1}
In~\cite{packard1988power}, it was shown that if $2s + f \leq 3$, then the right-hand side of Theorem~\ref{thm:bounds_mu} becomes an equality. To construct a fair and meaningful benchmark for our method based on this insight, we have constructed a specific random $3 \times 3$ MIMO system $\ve{G}_0(z)$ with $\ve{\Delta} \subset \mathbb{C}^{3 \times 3}$ such that the lower and upper bounds of $\mu_\Delta(\ve{G}_0)$ coincide. This guarantees that the true value of $\mu_\Delta$ is known, enabling a direct and reliable comparison between our estimate and the output of \texttt{mussv}. Eight distinct block structures are considered, as described in Table~\ref{tab:uncertainty_cases}, and evaluated in terms of (i) the convergence of $\tilde{\mu}$ and $\bar{\mu}$, and (ii) the number of frequency samples required to reach this convergence. Although not all block configurations satisfy the condition $2s + f \leq 3$ (e.g., when $s \geq 2$), we have identified a specific system $\ve{G}_0(z)$ for which this condition holds across all uncertainty structures considered.


Extensive simulations have demonstrated that, in most cases where $s=0$ and $f=1$ (i.e., a single full block for a given $n$, with $\ve{M} \in \mathbb{C}^{n \times n}$), both $\tilde{\mu}$ and $\bar{\mu}$ converge, and their values match the lower bound provided by \texttt{mussv}. Similarly, when the number of full blocks exceeds the number of scalar blocks ($f > s$), the algorithm generally exhibits good performance (and often when $m_k>r_j$ for all $j$, $k$). An exception to these cases occurs when the uncertainty structure satisfies $s=n$ (a single $n \times n$ scalar block) and $f=0$, in which case $\tilde{\mu} \neq \bar{\mu}$ for most instances, indicating poorer convergence.




\begin{table}[h!]
    \centering
    \caption{Uncertainty structures $\ve{\Delta} \in \mathbb{C}^{3\times3}$.}
    \label{tab:uncertainty_cases}
    \begin{tabular}{c|c|c}
        \hline
        Case & $\ve{r}$ (Scalar) & $\ve{m}$ (Full) \\
        \hline
        1 & $[1,3]$ & $[0,0]$ \\  
        2 & $[0,0]$ & $[1,3]$ \\  
        3 & $[1,2]$ & $[1,1]$ \\  
        4 & $[1,1]$ & $[1,2]$ \\  
        5 & $[2,1,1]$ & $[1,1]$ \\  
        6 & $[1,1]$ & $[2,1,1]$ \\  
        7 & $[3,1,1,1]$ & $[0,0]$ \\  
        8 & $[0,0]$ & $[3,1,1,1]$ \\  
        \hline
    \end{tabular}
\end{table}

\begin{figure}[h!]
    \centering
    \vspace{0.4 em}
\includegraphics[width=0.9\linewidth]{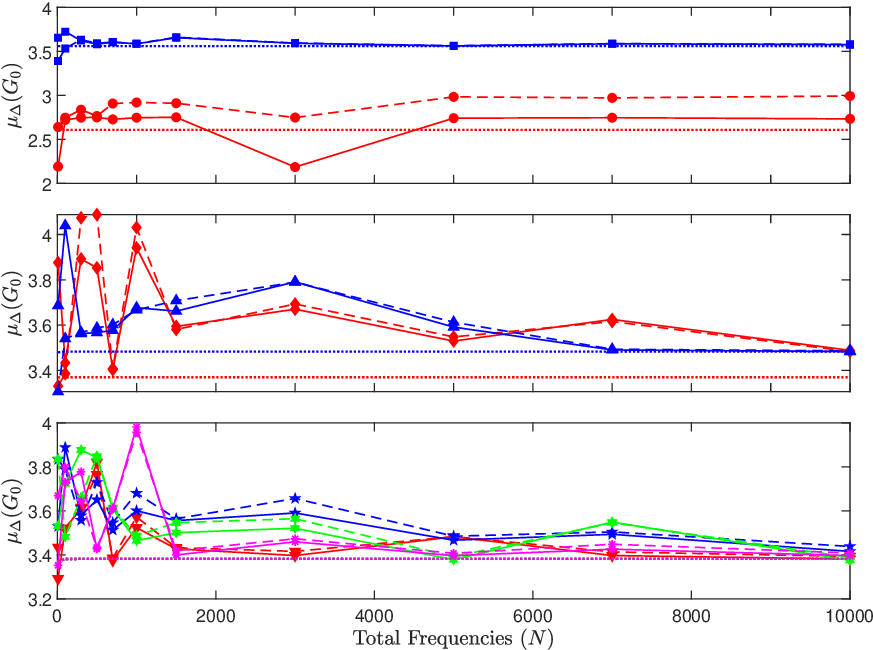} \\
    \vspace{1mm} 
    \includegraphics[width=0.7\linewidth]{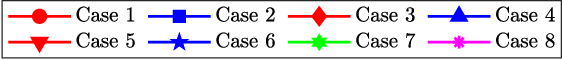}
\caption{Comparison between $\mu_{\Delta}(\ve{G}_0)$ and $N$. 
Solid line: $\tilde{\mu}$, dashed line: $\bar{\mu}$, dotted line: \texttt{mussv}. 
From top to bottom, the plots correspond to systems with 1 block, 2 blocks, and 3 blocks, respectively.}
    \label{fig:cases_legend}
    \vspace{-1.5 em}
\end{figure}

Figure~\ref{fig:cases_legend} presents the results for the configurations described in Table~\ref{tab:uncertainty_cases}. Note that, as $N$ increases, the performance of Algorithm~\ref{alg:power_method} improves, as previously discussed. Cases~1 ($f=0$) and~3 ($r_j>m_k$) do not converge to $\mu_M$; this behavior—involving large repeated scalar blocks—aligns with prior observations in~\cite{packard1988power, cheng1994continuous}, which suggest that power iteration methods may fail to converge or enter limit cycles under such structures. This is likely due to the limited directional excitation offered by scalar blocks, in contrast to full blocks that span richer subspaces and more effectively align with directions of maximal growth. Additionally, Cases~2 and 8 exhibit the best convergence across all cases, which is both encouraging and relevant for practical applications, as it suggests that the proposed algorithm is especially effective in computing reliable lower bounds in realistic robustness scenarios, where full complex blocks are commonly used to model input-output interactions and coupled perturbations.

\begin{figure*}[h!]
\begin{minipage}{0.32\textwidth}
    \centering
    \vspace{0.85 em}
\includegraphics[width=1\columnwidth]{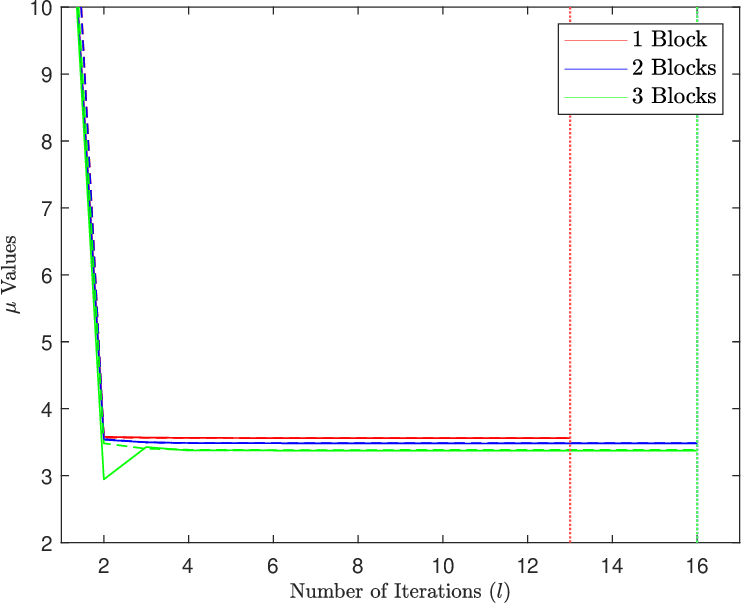}
    \caption{Convergence of the lower bound over iterations $(l)$. Solid line: $\tilde{\mu}$, dashed line: $\bar{\mu}$, dotted line: last iteration.}
    \label{fig:iteration}
\end{minipage}
\hfill 
\begin{minipage}{0.32\textwidth}
    \centering
\includegraphics[width=1\columnwidth]{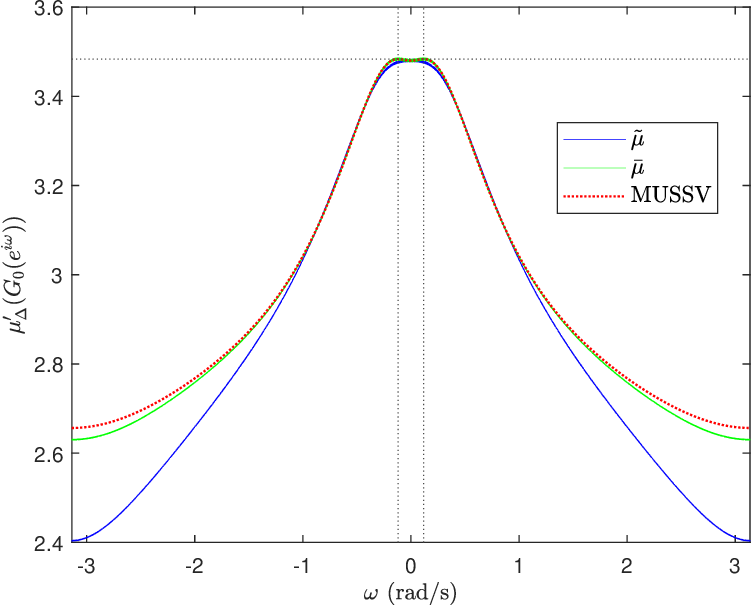}
    \caption{Frequency response of the structured singular value.}
    \label{fig:mu_plot}
\end{minipage}
\hfill 
\begin{minipage}{0.32\textwidth}
    \centering
\includegraphics[width=1\columnwidth]{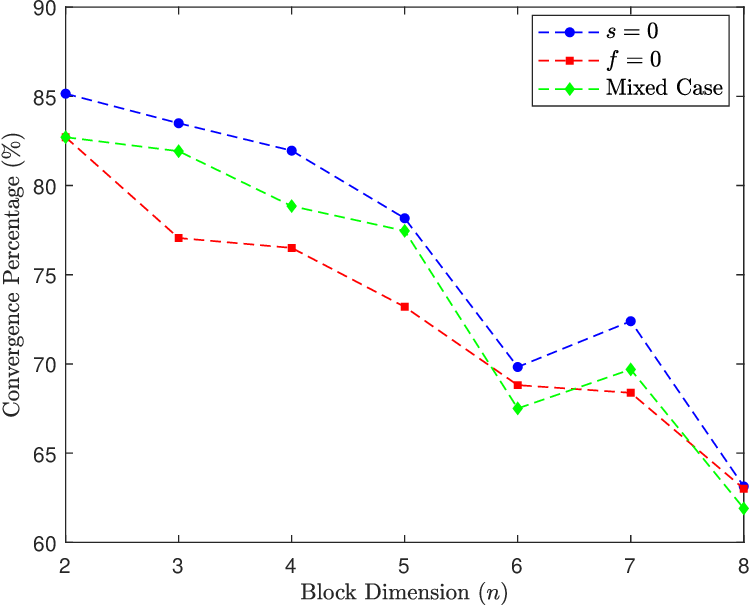}
    \caption{Percentage of converging cases of the data-driven power method.}
    \label{fig:error_plot}
\end{minipage}
\vspace{-1.5 em}
\end{figure*}

In practice, when the number of frequencies $N$ is large, the number of iterations $l$ required for the algorithm to converge typically ranges between 15 and 30. For the test under consideration, where $N=10000$ was used to ensure convergence, Figure~\ref{fig:iteration} illustrates the iterative behavior for cases where the lower bound successfully reaches \texttt{mussv}, highlighting the effect of the number of blocks on the stopping criterion, where a higher number of blocks requires more iterations to meet the termination condition. Only the converging cases are shown, thus excluding Cases~1 and 3. Cases~5–8, which all involve three blocks, required the same number of iterations.

Figure~\ref{fig:mu_plot} illustrates the structured singular value as a function of frequency for Case 4. The observed behavior, where $ \tilde{\mu} $ and $ \bar{\mu} $ align well with \texttt{mussv} at the dominant frequency but deviate at others, is expected based on the properties of the power method~\cite{rojas2012analyzing}. Indeed, the power method generates input signals whose energy iteratively concentrates around the frequency $\omega$ at which the lower bound on $\mu'_\Delta(\ve{G}_0(e^{i \omega}))$ is largest, which leads to better estimates at that frequency at the expense of poorer estimates for other frequencies.


Table~\ref{tab:ssv_noise} illustrates how Gaussian noise in the output, with variance $\sigma^2$, affects the convergence of $\tilde{\mu}$ and $\bar{\mu}$, which show that the algorithm can be sensitive to noise. The algorithm's robustness to noise can be improved, using, \eg, instrumental variables methods, but this is left for future research.

\begin{table}[h!]
    \centering
    \caption{Structured singular values, and peak frequencies, for different noise levels.}
    \label{tab:ssv_noise}
    \begin{tabular}{c|cc|cc}
        \hline
        \multirow{2}{*}{$\sigma^2$} & \multicolumn{2}{c|}{$\tilde{\mu}$} & \multicolumn{2}{c}{$\bar{\mu}$} \\
        & Value & Frequency & Value & Frequency \\
        \hline
        0      & 3.4832  & 0.0000  & 3.4864  & 0.1137 \\
        $10^{-6}$ & 3.9034  & 0.0415  & 4.1362  & 0.0258 \\
        $10^{-5}$ & 4.7400  & 0.1005  & 6.0842  & 0.4876 \\
        $10^{-4}$ & 8.2426  & 2.9016  & 11.6509 & 0.3914 \\
        $10^{-3}$ & 22.4976 & 2.3989  & 33.6600 & 0.4115 \\
        0.01   & 65.4947 & 1.6757  & 125.8755 & 2.7307 \\
        \hline
        \verb|MUSSV|  & 3.4833  & 0.1181  & --      & -- \\
        \hline
    \end{tabular} 
\end{table}
In summary, this benchmark confirms that Algorithm~\ref{alg:power_method} effectively estimates the structured singular value under favorable structural conditions, while also highlighting directions for improvement in noisy or scalar-dominated scenarios.

\newpage
\subsection*{ Test $\#$2}
To obtain more general insights into the convergence behavior of Algorithm~\ref{alg:power_method}, we have conducted a large set of simulations using randomly generated $n \times n$ complex systems, with $n = 2, 3, \dots, 8$. For each $n$, we tested the algorithm under the following three block structure configurations: (i) $\ve{r}=[0,0]$ and $\ve{m}=[1,n]$, (ii) $\ve{r}=[n,1,\cdots,1]$ and $\ve{m}=[0,0]$, (iii) for even $n$, $\ve{r}=[n/2,1,\dots,1]$ and $\ve{m}=[1,n/2]$; for odd $n$, $\ve{r}=[(n-1)/2,1,\dots,1]$ and $\ve{m}=[1,(n+1)/2]$. We ran 700 experiments per configuration, totaling 2100 simulations.

Figure~\ref{fig:error_plot} illustrates the percentage of simulations where the average of $\tilde{\mu}$ and $\bar{\mu}$ converge to $\mu_M$, which demonstrates better performance for $s=0$ and, across all three cases, a deterioration of the algorithm when $n>5$. It should be noted that, for $s + 2f > 3$, the exact value of $\mu_{\Delta}(\ve{G}_0)$ is unknown, potentially introducing bias when comparing the lower bounds to \texttt{mussv}. Nevertheless, using randomly generated synthetic systems provides a first step in evaluating algorithm performance, and future work could explore its applicability to real-world system dynamics. 

Overall, the results show that Algorithm~\ref{alg:power_method} yields reliable lower bound estimates for $\mu_\Delta(\ve{G}_0)$, especially with full blocks and moderate system order.
Additionally, it shows improved convergence in systems with faster dynamics, likely because shorter transients reduce the impact of non-steady-state data.



\section{Conclusion} \label{sec: conclusion}
In this paper, we have introduced a data-driven method to estimate a lower bound for the structured singular value of a dynamical system from input-output data. Our approach is model-free: a model of the system is not required nor explicitly built. Numerical examples show our method closely matches the lower bound from \verb|mussv|. For future work, we consider estimating a corresponding upper bound for the structured singular value from input-output data. 


 
\bibliographystyle{IEEEtran}
\bibliography{References}

\end{document}